\documentclass{elsart}

\usepackage{amssymb,amsmath,amstext}

\begin{document}

\begin{frontmatter}



\title{Second order gauge theory}


\author{R. R. Cuzinatto},
\ead{rodrigo@ift.unesp.br}
\author{C. A. M. de Melo}
\ead{cassius.anderson@gmail.com}
\address{Instituto de F\'{\i}sica Te\'{o}rica, Universidade Estadual
Paulista.\\
Rua Pamplona 145, CEP 01405-900, S\~{a}o Paulo, SP, Brazil.} and
\author{P. J. Pompeia}
\ead{pompeia@ift.unesp.br}
\address{Instituto de F\'{\i}sica Te\'{o}rica, Universidade Estadual
Paulista.\\
Rua Pamplona 145, CEP 01405-900, S\~{a}o Paulo, SP, Brazil.}
\address{Comando-Geral de Tecnologia Aeroespacial, Instituto de Fomento e Coordena\c{c}\~{a}o Industrial.
Pra\c{c}a Mal. Eduardo Gomes 50, CEP 12228-901, S\~{a}o Jos\'{e} dos
Campos, SP, Brazil.}


\begin{abstract}
A gauge theory of second order in the derivatives of the auxiliary
field is constructed following Utiyama's program. A novel field
strength $G=\partial F+fAF$\ arises besides the one of the first
order treatment, $F=\partial A-\partial A+fAA$. The associated
conserved current is obtained. It has a new feature: topological
terms are determined from local invariance requirements. Podolsky
Generalized Eletrodynamics is derived as a particular case in which
the Lagrangian of the gauge field is $L_{P}\propto G^{2}$. In this
application the photon mass is estimated. The $SU\left(  N\right)  $
infrared regime is analysed by means of Alekseev-Arbuzov-Baikov's
Lagrangian.
\end{abstract}

\end{frontmatter}

\section{Introduction\label{secao-Intro}}

Symmetry is one of the most important concepts in physics. Group
theoretical methods are used to find complete solutions for many
systems, particles were found by symmetry arguments, from Noether's
work we learn how to relate symmetries to conservation laws. In this
context a natural question is: by imposing to a system a generalized
version of some previous existent symmetry, would it be possible to
determine new features? In particular, starting from a kinetic
Lagrangian with a global symmetry, would it be possible generate
interactions via localization of the symmetry? In other words, could
local kinematical symmetries also imply dynamics?

The general answer to these questions is yes, if the local symmetry
is represented by a general Lie group, and it was found by Utiyama
\cite{Utiyama} in 1956 showing that it is possible to keep the
action invariant under point dependent transformations of the matter
field if one introduces a new field, the gauge potential,\textit{\
}and derives a minimal coupling prescription which determines the
fundamental interactions. Some years later, Ogievetski and
Polubarinov \cite{O-P} presented a criticism to this \emph{gauge
principle} by means of what is known today as B-Field formalism
\cite{NakaLivro,Bfield}. Notwithstanding, gauge principle remains as
a cornerstone of modern physics.

On the other hand, if one assumes that the equations of motion are
derivable from some Lagrangian, a natural way to generalize a given
theory is to suppose that the Lagrangian of the field contains terms
involving derivatives of higher order. Poincar\'{e} in 1901, when
discussing the law of inertia, already called our attention to the
importance of higher order equations \cite{Poincare}:
\textit{\textquotedblleft(...) The law of inertia (...) is not
imposed on us }\emph{a priori }\textit{(...). If a body is not acted
upon by a force, instead of supposing that its velocity is unchanged
we may suppose that its position or its acceleration is unchanged.
(...) in the second case, [we may suppose] that the variation of the
acceleration of a body depends only on the position of the body and
of neighbouring bodies, on their velocities and accelerations; or,
in mathematical terms, the differential equations of motion would be
of (...) third order (...).\textquotedblright} The non-singular
Hamilton-Lagrange theory was extended to arbitrary order by
Ostrogradsky \cite{Ostrogradsky} in 1850, generalizing the form of
canonical momenta. Following this reasoning, Bopp \cite{Bopp} and
Podolsky \cite{Podolsky} independently proposed a generalization of
electrodynamics containing second order derivatives. Quantization of
the theory resulted in finite energies in 1-loop approximation. This
leads to the idea that some quantum field corrections would be
simulated by the Podolsky's effective term. This generalized
electrodynamics was able also to explain the 4/3 factor in
Abrahaam-Lorentz theory \cite{Frenkel} as an electromagnetic mass
term by means of an appropriate gauge choice and a Lorentz invariant
cutoff induced by quantum effects. This cutoff provides an natural
mechanism to estimate the photon mass in the Podolsky generalized
electrodynamics, as we will show in the section \ref{MassPhoton} --
the presence of this massive mode does not violate the gauge
invariance as it occurs in the first order approach, and it is an
intrinsic feature of the second order gauge theory. Inspired by
Podolsky's work, Green \cite{Green} included another term involving
the second derivative obtaining a generalized meson-field theory
with finite energies at 1-loop. The relative success of these
achievements motivated some authors to propose finite extensions of
Quantum Chromodynamics (QCD) \cite{Confinamento} and also to
advocate that higher order terms would be able to explain the quark
confinement. The undesired feature of the Podolsky's theory is the
lack of unitarity at 1-loop for its quantum version, and it is an
open question if this characteristic can be ruled out in a
nonperturbative scheme. Besides all these motivations we emphasize
that, from a theoretical point of view, higher order theories have
many interesting features that justify their study by itself.

The scope of the present work is to construct a gauge theory for
Lagrangians of second order on the auxiliary field, following
Utiyama's procedure. Section \ref{secao-Utiyama}\ is\ devoted to
fixing the notation, to reviewing how the gauge potential $A$\
appears and why it ensures the invariance of the action integral. In
the next section, we assume a Lagrangian of the type $L\left(
A,\partial A,\partial^{2}A\right)  $\ for the gauge potential and
show that this Lagrangian must depend on $A$ and its derivatives
only through the usual field strength $F=\partial A-\partial A+fAA$\
and a second field strength $G=\partial F+fAF$, a new quantity that
naturally turns up\ in the second order theory. The covariance of
$F$\ and $G $\ under the gauge transformation is proved in the same
section \ref{secao-SegOrd}\ and the Bianchi identities for these two
objects are also deduced.

After settling\ the basis of the general theory, we proceed to the
analysis of the current derived from the total Lagrangian -- the
matter one plus the gauge Lagrangian\ -- in section
\ref{secao-Corrente}. If Utiyama's definition of current is kept,
one obtains a \emph{quasi}conserved current instead of a conserved
one. An alternative choice is to define the current in a different
fashion to enforce conservation.

As an application, in\ section \ref{secao-U(1)}, it is demonstrated
that the Podolsky Generalized Electrodynamics can be derived from
the second order gauge theory from a Lagrangian of the type
$L_{P}\propto G^{2}$. Section \ref{AlekseevArbuzov}\ deals with a
particular non-abelian case.


\section{Local Invariance and the Gauge Field\label{secao-Utiyama}}

Let $Q^{A}(x),$ $(A=1,2,...,N)$ be a general matter field whose
Lagrangian
density is%
\[
L(Q^{A},\partial_{\mu}Q^{A}),\text{ \ \ \ \ \ }\partial_{\mu}Q^{A}%
=\frac{\partial Q^{A}}{\partial x^{\mu}},
\]
with equation of motion given by%
\begin{equation}
\frac{\partial L}{\partial Q^{A}}-\partial_{\mu}\frac{\partial L}%
{\partial\left(  \partial_{\mu}Q^{A}\right)  }=0. \label{eq de campo}%
\end{equation}
We postulate the action integral,
\[
I=\int_{\Omega}L\,d^{4}x\,,
\]
to be invariant under the global infinitesimal transformation:%
\begin{align}
Q^{A}  &  \rightarrow Q^{A}+\delta Q^{A},\nonumber\\
\delta Q^{A}  &  =T_{a\text{ }B}^{\text{ }A}\epsilon^{a}Q^{B}, \label{1.1}%
\end{align}
where $\epsilon^{a}$ is an infinitesimal parameter independent of
$x$ $(a=1,2,...,n)$ and $T_{a\text{ }B}^{\text{\ }A}$ are constant
matrices.\ In what follows, we assume that the transformation
(\ref{1.1}) belongs to a Lie group $G$ dependent on $n$ parameters
$\epsilon^{a}$. Then the structure
constants $f_{a\text{ }c}^{\text{ }b}$\ are defined by%
\begin{equation}
\lbrack T_{a},T_{b}]_{\text{ }B}^{A}=f_{a\text{ }b}^{\text{ }c}T_{c\text{ }%
B}^{\text{\ }A}, \label{1.2}%
\end{equation}
satisfying
\begin{equation}
f_{a\text{ }b}^{\text{ }m}f_{m\text{ }c}^{\text{ }l}+f_{b\text{
}c}^{\text{ }m}f_{m\text{ }a}^{\text{ }l}+f_{c\text{ }a}^{\text{
}m}f_{m\text{ }b}^{\text{
}l}=0, \label{1.3 a}%
\end{equation}
which is the same as Jacobi identity, and%
\begin{equation}
f_{a\text{ }b}^{\text{ }c}=-f_{b\text{ }a}^{\text{ }c}, \label{1.3 b}%
\end{equation}
in accordance with (\ref{1.2}).

From the invariance of the action under (\ref{1.1}) in any spacetime
volume
$\Omega$, it results%
\begin{equation}
\delta L\equiv\frac{\partial L}{\partial Q^{A}}\delta
Q^{A}+\frac{\partial L}{\partial\left(  \partial_{\mu}Q^{A}\right)
}\delta\left(  \partial_{\mu
}Q^{A}\right)  \equiv0. \label{1.4}%
\end{equation}
Using the independence of the parameters we find,%
\begin{equation}
\frac{\partial L}{\partial Q^{A}}T_{a\text{ }B}^{\text{ \ \ }A}Q^{B}%
+\frac{\partial L}{\partial\left(  \partial_{\mu}Q^{A}\right)  }T_{a\text{ }%
B}^{\text{ }A}\partial_{\mu}Q^{B}\equiv0. \label{1.5}%
\end{equation}
In order to write (\ref{1.5}) we consider a variation $\delta$ which
is strictly functional, i.e., the spacetime point is not changed.

Rewriting (\ref{1.4}) by using Leibniz rule,%
\[
\left\{  \frac{\partial L}{\partial
Q^{A}}-\partial_{\mu}\frac{\partial L}{\partial\left(
\partial_{\mu}Q^{A}\right)  }\right\}  \delta
Q^{A}+\partial_{\mu}\left(  \frac{\partial L}{\partial\left(
\partial_{\mu }Q^{A}\right)  }\delta Q^{A}\right)  =0.
\]
The field equation (\ref{eq de campo}) must be satisfied, so that%
\begin{equation}
\partial_{\mu}J_{\text{ }a}^{\mu}=0,\text{ \ \ }J_{\text{ }a}^{\mu}\equiv
\frac{\partial L}{\partial\left(  \partial_{\mu}Q^{A}\right)  }T_{a\text{ }%
B}^{\text{\ }A}Q^{B}. \label{1.6}%
\end{equation}

Now let us consider the following infinitesimal transformation with
a \emph{point dependent} parameter $\epsilon^{a}(x)$
($a=1,2,...,n$):%
\begin{align}
&\delta Q^{A}(x)=T_{a\text{ }B}^{\text{\ }A}\epsilon^{a}(x)Q^{B},\nonumber\\
&T_{a\text{ }B}^{\text{\ }A}=\text{\textit{constant coefficients}%
},\label{1.1 linha}
\end{align}
In this case,%
\begin{align}
\delta L  \equiv&\left\{  \frac{\partial L}{\partial Q^{A}}T_{a\text{ }%
B}^{\text{ \ \ }A}Q^{B}+\frac{\partial L}{\partial\left(  \partial_{\mu}%
Q^{A}\right)  }T_{a\text{ }B}^{\text{ }A}\partial_{\mu}Q^{B}\right\}
\epsilon^{a}(x)+\nonumber\\&+\frac{\partial L}{\partial\left(
\partial_{\mu}Q^{A}\right) }T_{a\text{ }B}^{\text{\
}A}Q^{B}\partial_{\mu}\epsilon^{a}(x),
\end{align}
or%
\begin{equation}
\delta L\equiv\frac{\partial L}{\partial\left(
\partial_{\mu}Q^{A}\right) }T_{a\text{ }B}^{\text{\
}A}Q^{B}\partial_{\mu}\epsilon^{a}(x).
\label{1.5 linha}%
\end{equation}
We see in this case that $\delta L$\ does not vanish -- still
assuming that (\ref{1.5}) is valid even when
$\epsilon^{a}=\epsilon^{a}(x)$.

If one wants to preserve invariance of the Lagrangian under
(\ref{1.1 linha}), it is necessary to introduce a new field
\cite{Utiyama}, called \emph{gauge potential}, $A_{\,\mu}^{a}\left(
x\right)$\, which transforms as%
\begin{equation}
\delta
A_{\,\mu}^{c}=f_{a\,b}^{\,c}A_{\,\mu}^{b}\epsilon^{a}(x)+\partial_{\mu
}\epsilon^{c}, \label{DeltaA}%
\end{equation}
appearing in a new Lagrangian
$L^{\prime}(Q^{A},\partial_{\mu}Q^{A},A_{\,\mu
}^{a})$ through the combination%
\begin{equation}
\nabla_{\mu}Q^{A}\equiv\partial_{\mu}Q^{A}-T_{a\text{ }B}^{\text{\ }A}%
Q^{B}A_{\,\mu}^{a}. \label{1.10}%
\end{equation}

This new object is covariant, since it transforms exactly as the
original
field, i.e.,%
\[
\delta\left(  \nabla_{\mu}Q^{A}\right)  =T_{a\text{ }B}^{\text{\ }A}%
\nabla_{\mu}Q^{B}\epsilon^{a}(x),
\]
and it substitutes the ordinary derivative in the original
Lagrangian, in a prescription named\textit{\ }\emph{minimal
coupling} for description of the
interaction,%
\[
L^{\prime}(Q^{A},\partial_{\mu}Q^{A},A_{\,\mu}^{a})=L(Q^{A},\nabla_{\mu}%
Q^{A}).
\]
Usual derivatives are replaced, in the original Lagrangian, by the
\emph{covariant derivative} given by (\ref{1.10}).


\section{Gauge Field Second Order Lagrangian\label{secao-SegOrd}}

Let us take a Lagrangian for the auxiliary field as a functional
kernel
dependent up to second order derivatives,%
\[
L_{0}(A_{\text{ }\mu}^{a},\partial_{\nu}A_{\text{
}\mu}^{a},\partial_{\alpha }\partial_{\nu}A_{\text{
}\mu}^{a}),\text{ \ \ \ \ }\partial_{\nu}A_{\text{
}\mu}^{a}\equiv\frac{\partial A_{\ \mu}^{a}}{\partial x^{\nu}}.
\]
We will impose that this kernel is invariant under (\ref{DeltaA}), so%
\begin{gather*}
\delta L_{0}\equiv\frac{\partial L_{0}}{\partial A_{\text{
}\mu}^{a}}\delta A_{\text{ }\mu}^{a}+\frac{\partial
L_{0}}{\partial\left(  \partial_{\nu
}A_{\text{ }\mu}^{a}\right)  }\delta\left(  \partial_{\nu}A_{\text{ }\mu}%
^{a}\right)  +\\
+\frac{\partial L_{0}}{\partial\left(
\partial_{\alpha}\partial_{\nu }A_{\text{ }\mu}^{a}\right)
}\delta\left(  \partial_{\alpha}\partial_{\nu }A_{\text{
}\mu}^{a}\right)  \equiv0.
\end{gather*}
Substituting the transformation law for the gauge potential in this
equation and since $\epsilon^{a}$\ and its derivatives must be
functionally
independent, we are led to%
\begin{gather}
\frac{\partial L_{0}}{\partial A_{\text{ }\mu}^{a}}f_{c\text{ }b}^{\text{ }%
a}A_{\text{ }\mu}^{b}+\frac{\partial L_{0}}{\partial\left(
\partial_{\nu }A_{\text{ }\mu}^{a}\right)  }f_{c\text{ }b}^{\text{
}a}\partial_{\nu }A_{\text{ }\mu}^{b}+\frac{\partial
L_{0}}{\partial\left(
\partial_{\alpha}\partial_{\nu }A_{\text{ }\mu}^{a}\right)
}f_{c\text{ }b}^{\text{ }a}\partial_{\alpha
}\partial_{\nu}A_{\text{ }\mu}^{b}\equiv0, \label{1.15}%
\end{gather}%
\begin{gather}
\frac{\partial L_{0}}{\partial A_{\text{ }\nu}^{c}}+\frac{\partial L_{0}%
}{\partial\left(  \partial_{\nu}A_{\text{ }\mu}^{a}\right)  }f_{c\text{ }%
b}^{\text{ }a}A_{\text{ }\mu}^{b}+\left(  \frac{\partial
L_{0}}{\partial\left(
\partial_{\nu}\partial_{\alpha }A_{\text{ }\mu}^{a}\right)
}+\frac{\partial L_{0}}{\partial\left(
\partial_{\alpha}\partial_{\nu}A_{\text{ }\mu}^{a}\right)  }\right)
f_{c\text{ }b}^{\text{ }a}\partial_{\alpha}A_{\text{
}\mu}^{b}\equiv0,
\label{1.16}%
\end{gather}%
\begin{gather}
\frac{\partial L_{0}}{\partial\left(  \partial_{\nu}A_{\text{ }\alpha}%
^{c}\right)  }+\frac{\partial L_{0}}{\partial\left(
\partial_{\alpha
}A_{\text{ }\nu}^{c}\right)  }+\left(  \frac{\partial
L_{0}}{\partial\left(
\partial_{\nu}\partial_{\alpha }A_{\text{ }\mu}^{a}\right)
}+\frac{\partial L_{0}}{\partial\left(
\partial_{\alpha}\partial_{\nu}A_{\text{ }\mu}^{a}\right)  }\right)
f_{c\text{ }b}^{\text{ }a}A_{\text{ }\mu}^{b}\equiv0, \label{1.17}%
\end{gather}%
\begin{equation}
\frac{\partial L_{0}}{\partial\left(
\partial_{\alpha}\partial_{\nu}A_{\text{ }\mu}^{a}\right)
}+\frac{\partial L_{0}}{\partial\left(  \partial_{\nu
}\partial_{\mu}A_{\text{ }\alpha}^{a}\right)  }+\frac{\partial L_{0}}%
{\partial\left(  \partial_{\mu}\partial_{\alpha}A_{\text{
}\nu}^{a}\right)
}\equiv0. \label{1.17 linha}%
\end{equation}

This set of equations forms a hierarchy informing us about the
dependence of the Lagrangian with respect to the gauge potential and
its derivatives. To solve this hierarchy is the main aim of
Utiyama's general program. The solution of the above system gives
the covariant objects of the theory as well as the functional
dependence of the Lagrangian on these objects.

\subsection{Hierarchical Equations Solution}

\subsubsection{Solution of the Equation (\ref{1.17 linha})}

Once equation (\ref{1.17 linha}) involves only the dependence of the
Lagrangian density on the second derivatives of the gauge potential,
one may propose that this dependence appears through an object
$R_{\,\,\alpha\nu\mu
}^{a}$ which\ \emph{must} have a cyclic permutation symmetry:%
\begin{equation}
R_{\,\,\alpha\nu\mu}^{a}+R_{\,\,\nu\mu\alpha}^{a}+R_{\,\,\mu\alpha\nu}%
^{a}\equiv0. \label{ciclicidade de R}%
\end{equation}
The most general linear object constructed from $\partial^{2}A$ with
this
property is given by:%
\begin{equation}
R_{\,\,\alpha\nu\mu}^{a}=\partial_{\nu}\partial_{\alpha}A_{\text{ }\mu}%
^{a}-\partial_{\alpha}\partial_{\mu}A_{\text{ }\nu}^{a}, \label{R2}%
\end{equation}
where we chose $A\equiv1$ without loss of generality.

\subsubsection{Solution of the Equation (\ref{1.17})}

Let us consider now the equation (\ref{1.17}) written in terms of
$L_{0}^{\left(  1\right)  }\left(  A,\partial A,R\right)  $:%
\begin{gather*}
\frac{\partial L_{0}^{\left(  1\right)  }}{\partial\left(
\partial_{\nu }A_{\text{ }\alpha}^{c}\right)  }+\left[
\frac{\partial L_{0}^{\left(
1\right)  }}{\partial\left(  R_{\text{ }\nu\alpha\mu}^{a}\right)  }%
-\frac{\partial L_{0}^{\left(  1\right)  }}{\partial\left(  R_{\text{ }%
\alpha\mu\nu}^{a}\right)  }\right]  f_{c\text{ }b}^{\text{
}a}A_{\text{ }\mu
}^{b}+\\
+\frac{\partial L_{0}^{\left(  1\right)  }}{\partial\left(
\partial_{\alpha }A_{\text{ }\nu}^{c}\right)  }+\left[
\frac{\partial L_{0}^{\left(  1\right) }}{\partial\left(  R_{\text{
}\alpha\nu\mu}^{a}\right)  }-\frac{\partial
L_{0}^{\left(  1\right)  }}{\partial\left(  R_{\text{ }\nu\mu\alpha}%
^{a}\right)  }\right]  f_{c\text{ }b}^{\text{ }a}A_{\text{
}\mu}^{b}\equiv0.
\end{gather*}

This equation shows a symmetry in $\nu\leftrightarrow\alpha$, which
must be present in its solution. Also, the solution will be
construct from $R$ and then it must respect the cyclic symmetry as
dicted by (\ref{1.17 linha}). Therefore, the solution for this
functional partial differential equation is such that the functional
dependence of $L_{0}^{\left(  1\right)  }$\ with
respect to $R$\ shall be through the object%
\begin{equation}
Q_{\text{ }\beta\rho\sigma}^{d}\equiv R_{\text{ }\beta\rho\sigma}%
^{d}-f_{c\text{ }b}^{\text{ }d}\left[  A_{\text{
}\sigma}^{b}\partial_{\rho }A_{\text{ }\beta}^{c}-A_{\text{
}\rho}^{b}\partial_{\sigma}A_{\text{ }\beta
}^{c}\right]  . \label{Q2}%
\end{equation}
With this new quantity we pass from $L_{0}^{\left(  1\right)
}\left( A,\partial A,R\right)  $\ to $L_{0}^{\left(  2\right)
}\left(  A,\partial A,Q\right)  $.

A second independent solution to (\ref{1.17}) can be found if one
rewrites it
in terms of $L_{0}^{\left(  2\right)  }\left(  A,\partial A,Q\right)  $, i.e.,%
\[
\left[  \frac{\partial L_{0}^{\left(  2\right)  }}{\partial\left(
\partial_{\nu}A_{\text{ }\alpha}^{c}\right)  }+\frac{\partial L_{0}^{\left(
2\right)  }}{\partial\left(  \partial_{\alpha}A_{\text{
}\nu}^{c}\right) }\right]  _{Q}\equiv0.
\]
In this case, the dependence on the second derivative is eliminated
and the dependence on the first derivative can figure only
\textit{via}
the\ combination%
\[
A_{\left[  \nu\alpha\right]  }^{c}\equiv\partial_{\nu}A_{\text{ }\alpha}%
^{c}-\partial_{\alpha}A_{\text{ }\nu}^{c}.
\]
Thus, when we construct $Q$ we are actually selecting a sector of
the gauge potential Lagrangian. Here we have introduced the
antisymmetric operation $O_{\left[  \mu\nu\right]  }\equiv
O_{\mu\nu}-O_{\nu\mu}$ for a general $O_{\mu\nu}$. So, one goes from
$L_{0}^{\left(  2\right)  }\left(  A,\partial A,Q\right)  $\ to
\[
L_{0}^{\left(  3\right)  }=L_{0}^{\left(  3\right)  }\left(
A_{\,\mu}^{a},A_{ \left[  \nu\alpha\right]  }^{c},Q_{\text{
}\beta\rho\sigma}^{d}\right)  .
\]

\subsubsection{Solution of the Equation (\ref{1.16})}

Our next step is to rewrite the equation (\ref{1.16}) in terms of
the new functional dependence of the Lagrangian,
$L_{0}=L_{0}^{\left(  3\right)
}\left(  A,A_{\left[  \,\right]  },Q\right)  $:%
\begin{align*}
&  \frac{\partial L_{0}^{\left(  3\right)  }}{\partial A_{\text{ }\nu}^{a}%
}+\frac{\partial L_{0}^{\left(  3\right)  }}{\partial A_{\left[
\rho \sigma\right]  }^{d}}\left(
\delta_{\,\rho}^{\nu}\delta_{\,\sigma}^{\alpha
}-\delta_{\,\sigma}^{\nu}\delta_{\,\rho}^{\alpha}\right)  f_{a\text{ }%
e}^{\text{ }d}A_{\text{ }\alpha}^{e}+\\
&  +\frac{\partial L_{0}^{\left(  3\right)  }}{\partial Q_{\text{
}\beta \rho\sigma}^{d}}\left[  \left(
\delta_{\sigma}^{\,\nu}A_{\,\left[  \rho \beta\right]
}^{c}+\delta_{\beta}^{\,\nu}A_{\,\left[  \rho\sigma\right]
}^{c}+\delta_{\rho}^{\,\nu}A_{\,\left[  \beta\sigma\right]
}^{c}\right) f_{a\text{ }c}^{\text{ }d}+A_{\text{ }\beta}^{e}\left(
A_{\text{ }\sigma}^{b}\delta_{\,\rho
}^{\nu}-A_{\text{ }\rho}^{b}\delta_{\,\sigma}^{\nu}\right)  f_{b\text{ }%
c}^{\text{ }d}f_{a\text{ }e}^{\text{ }c}\right]  \equiv0.
\end{align*}

The solution of this equation utters that the functional form of the
Lagrangian density depends on three objects:
\begin{equation}
F_{\,\rho\sigma}^{d}\equiv A_{\left[  \rho\sigma\right]  }^{d}+f_{a\text{ }%
e}^{\text{ }d}A_{\text{ }\rho}^{e}A_{\text{ }\sigma}^{a}, \label{F}%
\end{equation}
\begin{align}
G_{\text{ }\beta\rho\sigma}^{d}  &  \equiv Q_{\text{ }\beta\rho\sigma}%
^{d}-\left\{  \left(  \delta_{\sigma}^{\,\lambda}A_{\,\left[
\rho\beta\right]
}^{c}+\delta_{\beta}^{\,\lambda}A_{\,\left[  \rho\sigma\right]  }^{c}%
+\delta_{\rho}^{\,\lambda}A_{\,\left[  \beta\sigma\right]
}^{c}\right)
f_{g\text{ }c}^{\text{ }d}A_{\text{ }\lambda}^{g}+\right. \nonumber\\
&  \left.  -\left(  A_{\text{
}\sigma}^{b}\delta_{\,\rho}^{\lambda}-A_{\text{
}\rho}^{b}\delta_{\,\sigma}^{\lambda}\right)  f_{b\text{ }c}^{\text{ }%
d}f_{g\text{ }e}^{\text{ }c}A_{\text{ }\lambda}^{g}A_{\text{ }\beta}%
^{e}\right\}  \label{G}%
\end{align}
and $A_{\text{ }\mu}^{a}$ itself.

Keeping in mind $L_{0}^{\left(  4\right)  }=L_{0}^{\left(  4\right)
}\left( F_{d}^{\rho\sigma},G_{\text{ }\beta\rho\sigma}^{d},A_{\text{
}\nu}^{a}\right)
$, we reexpress (\ref{1.16}):%
\[
\frac{\partial L_{0}^{\left(  4\right)  }\left(  F,G,A\right)
}{\partial A_{\text{ }\nu}^{a}}=0.
\]
Due to this, the Lagrangian kernel cannot be explicitly dependent on
the gauge potential $A$, which is the real reason why gauge fields
are massless. Therefore, the presence of massive terms is only
possible if the gauge symmetry is broken.

\subsubsection{Solution of the Equation (\ref{1.15}) -- Condition on the
Lagrangian}

After all that, we must put the equation (\ref{1.15}) in terms of
the new
objects $F$ and $G$,%
\begin{gather}
\frac{\partial L_{0}^{\left(  4\right)  }}{\partial F_{d}^{\rho\sigma}%
}f_{s\text{ }b}^{\text{ }h}\left(  \frac{\partial
F_{d}^{\rho\sigma}}{\partial
A_{\text{ }\nu}^{h}}A_{\text{ }\nu}^{b}+\frac{\partial F_{d}^{\rho\sigma}%
}{\partial\left(  \partial_{\nu}A_{\,\zeta}^{h}\right)
}\partial_{\nu
}A_{\text{ }\zeta}^{b}\right)  + \label{1.15Fim}\\
+\frac{\partial L_{0}^{\left(  4\right)  }}{\partial G_{\text{
}\beta \rho\sigma}^{d}}f_{s\text{ }b}^{\text{ }h}\left(
\frac{\partial G_{\text{
}\beta\rho\sigma}^{d}}{\partial A_{\text{ }\nu}^{h}}A_{\text{ }\nu}^{b}%
+\frac{\partial G_{\text{ }\beta\rho\sigma}^{d}}{\partial\left(
\partial _{\nu}A_{\,\zeta}^{h}\right)  }\partial_{\nu}A_{\text{
}\zeta}^{b}+\frac{\partial G_{\text{
}\beta\rho\sigma}^{d}}{\partial\left(
\partial_{\xi}\partial_{\nu}A_{\,\zeta}^{h}\right)  }\partial_{\xi}%
\partial_{\nu}A_{\text{ }\zeta}^{b}\right)  \equiv0,\nonumber%
\end{gather}
or, applying the Leibniz rule and the Jacobi identity,
\begin{equation}
\frac{\partial L_{0}^{\left(  4\right)  }}{\partial F_{\,\rho\sigma}^{d}%
}f_{c\,h}^{\,d}F_{\rho\sigma}^{h}+\frac{\partial L_{0}^{\left(  4\right)  }%
}{\partial G_{\text{ }\beta\rho\sigma}^{d}}f_{c\text{ }h}^{\text{
}d}G_{\text{
}\beta\rho\sigma}^{h}\equiv0. \label{1.15Linha}%
\end{equation}
This is a condition on the Lagrangian of the gauge potential which
\emph{must} be satisfied for the local symmetry to hold.

\subsection{New Expression for $G$ and the Bianchi Identity}

In order to reduce the number of independent objects in the
expressions above we will write $G$ in terms of $F$ and $A$. To do
so, let us consider initially
that, from (\ref{R2}) and (\ref{F}), we get%
\[
R_{\,\,\beta\rho\sigma}^{h}=\partial_{\beta}F_{\,\rho\sigma}^{h}+f_{e\text{
}b}^{\text{ }h}\partial_{\beta}\left(
A_{\,\rho}^{e}A_{\,\sigma}^{b}\right) .
\]

Using these informations to evaluate $G$ it results%
\begin{align*}
G_{\text{ }\beta\rho\sigma}^{h}  &  =\partial_{\beta}F_{\,\rho\sigma}%
^{h}+f_{c\text{ }b}^{\text{ }h}\left[  A_{\,\rho}^{c}\partial_{\beta
}A_{\,\sigma}^{b}+A_{\text{ }\rho}^{b}\partial_{\sigma}A_{\text{ }\beta}%
^{c}\right]  +\\
&  +f_{c\text{ }b}^{\text{ }h}\left(  A_{\text{
}\beta}^{b}A_{\,\left[ \rho\sigma\right]  }^{c}+A_{\text{
}\rho}^{b}A_{\,\left[  \beta\sigma\right]
}^{c}\right)-f_{g\text{ }b}^{\text{ }h}f_{c\text{ }f}^{\text{ }g}A_{\text{ }\beta}%
^{f}\left(  A_{\text{ }\sigma}^{b}A_{\text{ }\rho}^{c}-A_{\text{ }\rho}%
^{b}A_{\text{ }\sigma}^{c}\right)  ,
\end{align*}
that, with (\ref{F}) and the Jacobi identity, conduct us to%
\begin{equation}
G_{\text{ }\beta\rho\sigma}^{a}=\partial_{\beta}F_{\,\rho\sigma}%
^{a}+f_{c\text{ }e}^{\text{ }a}A_{\text{ }\beta}^{e}F_{\,\rho\sigma}%
^{c}=D_{\,c\beta}^{a}F_{\,\rho\sigma}^{c}, \label{Gnovo}%
\end{equation}
where $D_{\beta}$ is a kind of Fock-Ivanenko derivative \cite{Fock}\
\begin{equation}
D_{\,c\beta}^{a}\equiv\delta_{\,c}^{a}\partial_{\beta}-\omega_{c\,\beta}%
^{\,a}\,,\quad\omega_{c\,\beta}^{\,a}\equiv f_{e\,c}^{\text{
}a}A_{\text{
}\beta}^{e}. \label{Fock-IvanenkoG}%
\end{equation}

This kind of structure in terms of a Fock-Ivanenko-like derivative
is
exhibited also by $F$\ when written as%
\begin{equation}
F_{\,\rho\sigma}^{d}=D_{\,a\rho}^{d}A_{\text{ }\sigma}^{a}-D_{\,a\sigma}%
^{d}A_{\,\rho}^{a}\equiv D_{\,a[\rho}^{d}A_{\text{ }\sigma]}^{a},
\label{Fnovo}%
\end{equation}
in which%
\begin{equation}
D_{\,a\rho}^{d}=\delta_{\,a}^{d}\partial_{\rho}-\omega_{a\,\rho}^{\,d}%
;\quad\omega_{a\,\rho}^{\,d}\equiv\frac{1}{2}f_{e\text{ }a}^{\text{ }%
d}A_{\text{ }\rho}^{e}, \label{Fock-IvanencoF}%
\end{equation}
with a factor $1/2$\ in the spin connection $\omega_{a\,\rho}^{\,d}%
$,\ different from eq. (\ref{Fock-IvanenkoG}). This difference
refers to the fact that $A^{e}$ is a connection and not a vector
like $F^{d}$.

With this new expression for $G$ we immediately verify that it\ is
antissymetric in its last two spacetime indices. This is consistent
with the antissymetry property required by equation (\ref{1.17}). On
the other hand, equation (\ref{1.17 linha}) stablishes that a cyclic
property must be present in those objects containing the second
derivative of the gauge field. In fact, starting from (\ref{Gnovo})
and with the expression of$\ F_{d}^{\rho\sigma}$ and the Jacobi
identity we find
\begin{equation}
G_{\text{ }\beta\rho\sigma}^{a}+G_{\text{
}\rho\sigma\beta}^{a}+G_{\text{
}\sigma\beta\rho}^{a}=D_{\,c\beta}^{a}F_{\,\rho\sigma}^{c}+D_{\,c\rho}%
^{a}F_{\,\sigma\beta}^{c}+D_{\,c\sigma}^{a}F_{\,\beta\rho}^{c}=0,
\label{Bianchi}%
\end{equation}
the so called Bianchi identity.

Thereafter, the Bianchi identity (\ref{Bianchi}) is a natural
consequence of the local symmetry in the second order formalism as
dictated by the hierarchical equations (\ref{1.15}) to (\ref{1.17
linha}), and not just an \textit{a priori} equality constructed with
$F$.

This identity and the manifest skew-symmetry of $F_{\,\mu\nu}^{a}$
in its
spacetime indices, allow us to rewrite $G$ as%
\begin{equation}
G_{\text{
}\beta\rho\sigma}^{a}=D_{\,c[\sigma}^{a}F_{\,\rho]\beta}^{c},
\label{Gnovissimo}%
\end{equation}
in the same suggestive form as (\ref{Fnovo}) for $F$. This kind of
structure will be explored below in the section \ref{HighOrderGen}.

\subsection{Transformation Laws}

Next, we construct the transformation law for $G$. Remembering
\cite{Utiyama},
we have%
\[
\delta A_{\text{ }\mu}^{a}=f_{c\text{ }b}^{\text{ }a}A_{\text{ }\mu}%
^{b}\epsilon^{c}(x)+\frac{\partial\epsilon^{a}}{\partial x^{\mu}}%
\]
and%
\begin{equation}
\delta F_{\text{ }\mu\nu}^{a}=\epsilon^{c}(x)f_{c\text{ }b}^{\text{ }%
a}F_{\text{ }\mu\nu}^{b}. \label{1.20}%
\end{equation}

Therefore%
\begin{align*}
\delta G_{\text{ }\beta\rho\sigma}^{a}  &  =\epsilon^{d}f_{d\text{
}g}^{\text{
}a}\partial_{\beta}F_{\text{ }\rho\sigma}^{g}+\epsilon^{d}f_{c\text{ }%
e}^{\text{ }a}f_{d\text{ }g}^{\text{ }c}A_{\text{ }\beta}^{e}F_{\text{ }%
\rho\sigma}^{g}+f_{d\text{ }g}^{\text{ }a}F_{\text{
}\rho\sigma}^{g}\partial_{\beta
}\epsilon^{d}+f_{c\text{ }e}^{\text{ }a}\delta A_{\text{ }\beta}^{e}%
F_{\,\rho\sigma}^{c}%
\end{align*}
and, applying the transformation law for the gauge potential and the
Jacobi
identity,%
\begin{equation}
\delta G_{\text{ }\beta\rho\sigma}^{a}=\epsilon^{d}(x)f_{d\text{
}c}^{\text{
}a}G_{\text{ }\beta\rho\sigma}^{c}. \label{1.20linha}%
\end{equation}

Then just like $F$, $G$ is also\emph{\ covariant} under the action
of the local Lie group,\footnote{We could have arrived at the same
conclusion with a
glance at (\ref{1.15Linha}), $\frac{\partial L_{0}^{\left(  4\right)  }%
}{\partial F_{\,\rho\sigma}^{d}}f_{c\,h}^{\,d}F_{\rho\sigma}^{h}%
+\frac{\partial L_{0}^{\left(  4\right)  }}{\partial G_{\text{
}\beta
\rho\sigma}^{d}}f_{c\text{ }h}^{\text{ }d}G_{\text{ }\beta\rho\sigma}%
^{h}\equiv0$, which is nothing but the invariance condition of the
Lagrangian $\delta L_{0}^{\left(  4\right)  }\equiv0$,
$\frac{\partial L_{0}^{\left(
4\right)  }}{\partial F_{\,\rho\sigma}^{d}}\delta F_{\,\rho\sigma}^{d}%
+\frac{\partial L_{0}^{\left(  4\right)  }}{\partial G_{\text{
}\beta \rho\sigma}^{d}}\delta G_{\text{
}\beta\rho\sigma}^{d}\equiv0$, and so, $\delta
F_{\,\rho\sigma}^{d}=\epsilon^{c}f_{c\,h}^{\,d}F_{\rho\sigma}^{h}$;
$\delta G_{\text{ }\beta\rho\sigma}^{d}=\epsilon^{c}f_{c\text{ }h}^{\text{ }%
d}G_{\text{ }\beta\rho\sigma}^{h}$, where we have used the fact that
the Lagrangian is a kernel functional of $F$\ and $G$\ alone.} i.e.,
it transforms like a vector under the action of the Lie group.


\section{Second Order Current \label{secao-Corrente}}

From the minimal coupling prescription the total Lagrangian is%
\[
L_{T}\left(  Q^{A},\nabla_{\mu}Q^{A},A,\partial
A,\partial^{2}A\right) =L\left(  Q^{A},\nabla_{\mu}Q^{A}\right)
+L_{0}\left(  F,G\right)  .
\]
Observe that the total Lagrangian is a function of the second
derivatives of $A$, by means of $G$ in $L_{0}$.

From the variational calculus,%
\[
\frac{\delta L}{\delta Q^{A}}=\frac{\partial L_{T}}{\partial Q^{A}}%
-\partial_{\nu}\frac{\partial L_{T}}{\partial\left(  \partial_{\nu}%
Q^{A}\right)  },
\]%
\[
\frac{\delta L_{T}}{\delta A_{~\mu}^{a}}=\frac{\partial
L_{T}}{\partial A_{~\mu}^{a}}-\partial_{\nu}\left(  \frac{\partial
L_{T}}{\partial\left(
\partial_{\nu}A_{~\mu}^{a}\right)  }\right)  +\partial_{\nu}\partial_{\lambda
}\left(  \frac{\partial L_{T}}{\partial\left(
\partial_{\nu}\partial _{\lambda}A_{~\mu}^{a}\right)  }\right)  .
\]

The vanishing of the variation of the total Lagrangian results in an
equation composed of two terms: a volumetric and a surface one,
which must be null independently, in a development similar to
Noether theorem.

\subsection{The Variation of the Total Lagrangian}

The variation of $L_{T}=L_{T}\left(  Q^{A},\partial Q^{A},A,\partial
A,\partial^{2}A\right)  $ is:%
\[
\delta L_{T}\equiv\left[  \frac{\delta L_{T}}{\delta A_{~\mu}^{a}}%
f_{c\,b}^{\,a}A_{\,\mu}^{b}\epsilon^{c}-\partial_{\mu}\left(
\frac{\delta
L_{T}}{\delta A_{~\mu}^{a}}\right)  \epsilon^{a}+\left(  \frac{\delta L_{T}%
}{\delta Q^{A}}\right)  \delta Q^{A}\right]  +\partial_{\nu}M^{\nu},
\]
where%
\begin{align*}
M^{\nu}  &  =\frac{\delta L_{T}}{\delta A_{~\nu}^{a}}\epsilon^{a}%
+\frac{\partial L_{T}}{\partial\left(  \nabla_{\nu}Q^{A}\right)
}\delta Q^{A}+\left[  \frac{\partial L_{0}^{\left(  4\right)
}}{\partial F_{\,\left[ \nu\mu\right]  }^{a}}+f_{a\text{ }b}^{\text{
}d}A_{\text{ }\rho}^{b}\left( \frac{\partial L_{0}^{\left(  4\right)
}}{\partial G_{\,\rho\left[  \nu \mu\right]  }^{d}}+\frac{\partial
L_{0}^{\left(  4\right)  }}{\partial
G_{\,\nu\left[  \rho\mu\right]  }^{d}}\right)  \right]  \delta A_{\,\mu}%
^{a}+\\
&  -\frac{1}{2}\partial_{\rho}\left[  \frac{\partial L_{0}^{\left(
4\right) }}{\partial G_{\,\nu\left[  \rho\mu\right]
}^{a}}+\frac{\partial
L_{0}^{\left(  4\right)  }}{\partial G_{\,\rho\left[  \nu\mu\right]  }^{a}%
}\right]  \delta A_{\,\mu}^{a}+\frac{1}{2}\left[  \frac{\partial
L_{0}^{\left(  4\right)  }}{\partial G_{\,\nu\left[  \rho\mu\right]  }^{a}%
}+\frac{\partial L_{0}^{\left(  4\right)  }}{\partial
G_{\,\rho\left[  \nu \mu\right]  }^{a}}\right]
\partial_{\rho}\left(  \delta A_{\,\mu}^{a}\right) ~,
\end{align*}
or%
\[
M^{\nu}\equiv N_{c}^{~\nu}\epsilon^{c}+O_{a}^{~\nu\mu}\partial_{\mu}%
\epsilon^{a}+\frac{1}{2}P_{a}^{~\rho\nu\mu}\partial_{\rho}\partial_{\mu
}\epsilon^{a}~,
\]
with%
\begin{align}
N_{c}^{~\nu}  &  \equiv\frac{\delta L_{T}}{\delta
A_{~\nu}^{c}}+\frac{\partial L_{T}}{\partial\left(
\nabla_{\nu}Q^{A}\right)  }T_{\left(  c\right)
~B}^{\;\;A}Q^{B}+\nonumber\\ & +\left[  \frac{\partial L_{0}^{\left(
4\right) }}{\partial F_{\,\left[  \nu\mu\right]  }^{a}}+f_{a\text{
}b}^{\text{ }d}A_{\text{ }\rho }^{b}\left(  \frac{\partial
L_{0}^{\left(  4\right)  }}{\partial G_{\,\rho\left[  \nu\mu\right]
}^{d}}+\frac{\partial L_{0}^{\left(  4\right) }}{\partial
G_{\,\nu\left[  \rho\mu\right]  }^{d}}\right)  \right]
f_{c\,e}^{\,a}A_{\,\mu}^{e}+\label{N}\\
&  -\frac{1}{2}\partial_{\rho}\left[  \frac{\partial L_{0}^{\left(
4\right) }}{\partial G_{\,\nu\left[  \rho\mu\right]
}^{a}}+\frac{\partial
L_{0}^{\left(  4\right)  }}{\partial G_{\,\rho\left[  \nu\mu\right]  }^{a}%
}\right]  f_{c\,e}^{\,a}A_{\,\mu}^{e}+\frac{1}{2}\left[
\frac{\partial
L_{0}^{\left(  4\right)  }}{\partial G_{\,\nu\left[  \rho\mu\right]  }^{a}%
}+\frac{\partial L_{0}^{\left(  4\right)  }}{\partial
G_{\,\rho\left[  \nu \mu\right]  }^{a}}\right]
f_{c\,b}^{\,a}\partial_{\rho}A_{\,\mu}^{b}~;
\nonumber%
\end{align}%
\begin{align}
O_{a}^{~\nu\mu}  &  \equiv\frac{\partial L_{0}^{\left(  4\right)
}}{\partial F_{\,\left[  \nu\mu\right]  }^{a}}+f_{a\text{
}b}^{\text{ }d}A_{\text{ }\rho }^{b}\left(  \frac{\partial
L_{0}^{\left(  4\right)  }}{\partial G_{\,\rho\left[  \nu\mu\right]
}^{d}}+\frac{\partial L_{0}^{\left(  4\right)
}}{\partial G_{\,\nu\left[  \rho\mu\right]  }^{d}}\right)  +\label{O}\\
&  -\frac{1}{2}\partial_{\rho}\left[  \frac{\partial L_{0}^{\left(
4\right) }}{\partial G_{\,\nu\left[  \rho\mu\right]
}^{a}}+\frac{\partial
L_{0}^{\left(  4\right)  }}{\partial G_{\,\rho\left[  \nu\mu\right]  }^{a}%
}\right]  +\frac{1}{2}\left[  \frac{\partial L_{0}^{\left(  4\right)  }%
}{\partial G_{\,\nu\left[  \mu\rho\right]  }^{c}}+\frac{\partial
L_{0}^{\left(  4\right)  }}{\partial G_{\,\mu\left[  \nu\rho\right]  }^{c}%
}\right]  f_{a\,b}^{\,c}A_{\,\rho}^{b}~;\nonumber
\end{align}
and%
\begin{equation}
P_{a}^{~\rho\nu\mu}\equiv\frac{1}{2}\left(  \frac{\partial
L_{0}^{\left( 4\right)  }}{\partial G_{\,\rho\left[  \nu\mu\right]
}^{a}}+\frac{\partial
L_{0}^{\left(  4\right)  }}{\partial G_{\,\mu\left[  \nu\rho\right]  }^{a}%
}\right)  ~. \label{P}%
\end{equation}

\subsection{Hierarchical Equations for the Current}

The vanishing of the volumetric term in $\delta L_{T}=0$\ gives the
equations
of motion%
\[
\frac{\delta L_{T}}{\delta Q^{A}}=0\,,\quad\frac{\delta
L_{T}}{\delta A_{~\mu }^{a}}=0.
\]

By the same way from the vanishing of the surface term,%
\begin{equation}
\partial_{\nu}M^{\nu}\equiv0, \label{DelM}%
\end{equation}
we have the following set of hierarchical equations:%
\begin{align}
\partial_{\nu}N_{c}^{~\nu}  &  \equiv0\label{condicao1}\\
N_{c}^{~\nu}+\partial_{\mu}O_{c}^{~\mu\nu}  &  \equiv0\label{condicao2}\\
O_{a}^{~\left(  \nu\mu\right)  }+\partial_{\rho}P_{a}^{~\nu\rho\mu}
&
\equiv0\label{condicao3}\\
P_{a}^{~\rho\nu\mu}+P_{a}^{~\nu\mu\rho}+P_{a}^{~\mu\rho\nu}  &
\equiv0
\label{condicao4}%
\end{align}

These equations constitute a hierarchical set of equations which
governs the conservation law associated to the local symmetry, in a
similar way that eqs. (\ref{1.15}) to (\ref{1.17 linha}) determined
the functional form of the Lagrangian $L_{0}$.

\subsection{Solution of the Hierarchical Equations for the Current}

Equations (\ref{condicao4}) and (\ref{condicao3}) are automatically
satisfied by virtue of the symmetries in $F_{\,\mu\nu}^{c}$\ and
$G_{\text{ }\mu\rho\nu }^{d}$.

Condition (\ref{condicao2}) will be used to deffine the current,
while the equation (\ref{condicao1}) sets the conservation law.

\subsubsection{\emph{Quasi}conserved Current}

Let us define, as done by Utiyama,%
\begin{align}
J_{c}^{\,\nu}  &  \equiv\frac{\partial L_{T}}{\partial A_{~\nu}^{c}}%
=-\frac{\partial L_{T}}{\partial\left(  \nabla_{\nu}Q^{A}\right)
}T_{\left( c\right)  ~B}^{\;\;A}Q^{B}-\frac{\partial L_{0}^{\left(
4\right)  }}{\partial F_{\,\left[  \nu\mu\right]
}^{a}}f_{c\,e}^{\,a}A_{\,\mu}^{e}-\frac{\partial
L_{0}^{\left(  4\right)  }}{\partial G_{\,\nu\rho\mu}^{d}}f_{c\,b}%
^{\,d}F_{\,\rho\mu}^{b}+\nonumber\\
&  +\left(  D_{~d\rho}^{a}\frac{\partial L_{0}^{\left(  4\right)
}}{\partial
G_{\,\rho\left[  \nu\mu\right]  }^{d}}\right)  f_{c\,b}^{\,a}A_{\,\mu}%
^{b}-\partial_{\rho}\left(  \frac{\partial L_{0}^{\left(  4\right)  }%
}{\partial G_{\,\rho\left[  \nu\mu\right]  }^{a}}f_{c\,b}^{\,a}A_{\,\mu}%
^{b}\right)  . \label{corrente}%
\end{align}
This definition is inspired by the most direct experimental sense of
current, as a measure of the response of the system under a
variation of the field.

The condition (\ref{condicao1}) now gives%
\begin{equation}
\partial_{\nu}J_{c}^{~\nu}=\partial_{\nu}\frac{\delta L_{T}}{\delta A_{~\nu
}^{c}}-\frac{1}{2}\partial_{\nu}\partial_{\rho}\left(  \left[
\frac{\partial
L_{0}^{\left(  4\right)  }}{\partial G_{\,\nu\left[  \rho\mu\right]  }^{d}%
}+\frac{\partial L_{0}^{\left(  4\right)  }}{\partial
G_{\,\rho\left[  \nu \mu\right]  }^{d}}\right]
f_{c\,b}^{\,d}A_{\,\mu}^{b}\right)  .
\label{QuasiConserv}%
\end{equation}

On the mass shell it follows the \emph{quasi}conservation of the
current $J_{a}^{\,\mu}$. The term
\textquotedblleft\emph{quasi}\textquotedblright\ is understood in
the sense that if one takes the integral of (\ref{QuasiConserv})
and chooses boundary conditions such that%
\[
\int_{\partial\Omega}d\sigma_{\nu}\partial_{\rho}\left(  \left[
\frac{\partial L_{0}^{\left(  4\right)  }}{\partial G_{\,\nu\left[
\rho \mu\right]  }^{d}}+\frac{\partial L_{0}^{\left(  4\right)
}}{\partial
G_{\,\rho\left[  \nu\mu\right]  }^{d}}\right]  f_{c\,b}^{\,d}A_{\,\mu}%
^{b}\right)  =0,
\]
the \emph{conservation} of $J_{a}^{\,\nu}$\ is globally recovered.
For instance, this kind of boundary condition occurs when
$A_{\,\rho}^{e}$\ and its first derivative are null on the boundary
$\partial\Omega$.

It is interesting to notice that in the case of an abelian group
this conservation is also achieved. The same occurs in the first
order approach, and that is because Utiyama's definition coincides
with Noether's current. This does not happen in the second order
theory, as one can immediately see from the non-conservation of
$J_{a}^{\,\nu}$. In order to establish a conserved current in the
second order approach, we must build another one based on Utiyama's
proposal, which is later compared with the standard Noether's
current. This is done in the following section.

\subsubsection{Conserved Current}

An alternative definition for the current is%
\begin{align}
\bar{J}_{c}^{\,\nu}  &  \equiv\frac{\partial L_{T}}{\partial A_{~\nu}^{c}%
}-\partial_{\mu}\frac{\partial L_{T}}{\partial\left(
\partial_{\mu}A_{~\nu
}^{c}\right)  }=\nonumber\\
&  =-\frac{\partial L_{T}}{\partial\left(  \nabla_{\nu}Q^{A}\right)
}T_{\left(  c\right)  ~B}^{\;\;A}Q^{B}-\frac{\partial L_{0}^{\left(
4\right)
}}{\partial F_{\,\left[  \nu\mu\right]  }^{a}}f_{c\,e}^{\,a}A_{\,\mu}%
^{e}-\frac{\partial L_{0}^{\left(  4\right)  }}{\partial G_{\,\nu\rho\mu}^{d}%
}f_{c\,b}^{\,d}F_{\,\rho\mu}^{b}+\nonumber\\
&  +\left(  D_{a~\rho}^{\;d}\frac{\partial L_{0}^{\left(  4\right)  }%
}{\partial G_{\,\rho\left[  \nu\mu\right]  }^{d}}\right)  f_{c\,b}%
^{\,a}A_{\,\mu}^{b}-\partial_{\mu}\left[  \frac{\partial
L_{0}^{\left(
4\right)  }}{\partial F_{\,\left[  \mu\nu\right]  }^{c}}+f_{c\text{ }%
b}^{\text{ }d}A_{\text{ }\rho}^{b}\frac{\partial L_{0}^{\left(  4\right)  }%
}{\partial G_{\,\rho\left[  \mu\nu\right]  }^{d}}\right]  . \label{corrente2}%
\end{align}

Therefore%
\[
\partial_{\nu}\bar{J}_{c}^{\,\nu}=\partial_{\nu}\left(  \frac{\delta L_{T}%
}{\delta A_{~\nu}^{c}}\right)  =0
\]
under the equations of motion, showing that the current
(\ref{corrente2}) is \emph{strictly} conserved.

\subsection{Concerning the Utiyama and Noether's Currents}

In the current (\ref{corrente}) the first two terms are the current
obtained by Utiyama in the first order formalism \cite{Utiyama}, the
third and fourth terms are of second order, and the last one is the
\emph{quasi}conservation term.

On the other hand the current (\ref{corrente2}) is composed by first
two terms of the first order current, plus second order terms
involving $G$, and the last is a topological term \cite{Felsager},
in the sense that it is conserved independently of the equations of
motion.

It is worth to remember that these topological terms cannot have
their origins explained by the dynamics, but the general local
invariance scheme brings them with it.

We can evaluate the transformation laws for (\ref{corrente}) and
(\ref{corrente2}):%
\begin{align*}
\delta J_{c}^{~\nu}= &
-\epsilon^{e}f_{e\,c}^{\,a}J_{a}^{~\nu}+\left[ \frac{\partial
L_{0}^{\left(  4\right)  }}{\partial F_{\,\left[  \nu \mu\right]
}^{a}}+f_{a\text{ }h}^{\text{ }d}A_{\text{ }\rho}^{h}\left(
\frac{\partial L_{0}^{\left(  4\right)  }}{\partial G_{\,\rho\left[
\nu \mu\right]  }^{d}}+\frac{\partial L_{0}^{\left( 4\right)
}}{\partial
G_{\,\mu\left[  \nu\rho\right]  }^{d}}\right)  \right]  f_{e\,c}^{\,a}%
\partial_{\mu}\epsilon^{e}+\\
& +\frac{\partial L_{0}^{\left(  4\right)  }}{\partial
G_{\,\rho\left[ \nu \mu\right]
}^{a}}f_{e\,c}^{\,a}\partial_{\rho}\partial_{\mu}\epsilon^{e}~,
\end{align*}%
\[
\delta\bar{J}_{c}^{~\nu}=-\epsilon^{e}f_{e\text{ }c}^{\text{ }a}\bar{J}%
_{a}^{~\nu}+\partial_{\mu}\left(  \frac{\partial L_{0}^{\left(  4\right)  }%
}{\partial G_{\,\rho\left[  \mu\nu\right]  }^{a}}+\frac{\partial
L_{0}^{\left(  4\right)  }}{\partial G_{\,\mu\left[  \rho\nu\right]  }^{a}%
}\right)  f_{e\,c}^{\,a}\partial_{\rho}\epsilon^{e}+\frac{\partial
L_{0}^{\left(  4\right)  }}{\partial G_{\,\rho\left[  \mu\nu\right]  }^{a}%
}f_{e\text{ }c}^{\text{
}a}\partial_{\mu}\partial_{\rho}\epsilon^{e}~.
\]
As expected from $J_{c}^{\,\nu}\equiv\frac{\partial L_{T}}{\partial
A_{~\nu }^{c}}$ and $\bar{J}_{c}^{\,\nu}\equiv\frac{\partial
L_{T}}{\partial A_{~\nu }^{c}}-\partial_{\mu}\frac{\partial
L_{T}}{\partial\left(  \partial_{\mu }A_{~\nu}^{c}\right)  }$ both
currents are not covariant. However, one can
define a \emph{covariant} current:%
\begin{equation}
j_{c}^{~\mu}\equiv\frac{\partial L\left(  Q,\nabla Q\right)
}{\partial A_{~\mu}^{c}}=-\frac{\partial L_{T}}{\partial\left(
\nabla_{\mu}Q^{A}\right)
}T_{c\text{ }B}^{\text{\ }A}Q^{B}, \label{CovCurrent}%
\end{equation}
which is not strictly conserved, but only \emph{covariantly} conserved:%
\begin{equation}
D_{c~\mu}^{~a}j_{a}^{~\mu}=0. \label{CovConservLaw}%
\end{equation}
We conclude that covariance and conservation never can be obtained
simultaneously; in order to maintain one we must sacrifice the
other.

To compare the proposed currents $J_{c}^{\,\nu}$\ and $\bar{J}_{c}^{\,\nu}%
$\ with Noether's one, we rewrite them as%
\begin{align*}
J_{c}^{~\nu} &  =-\frac{\partial L_{T}}{\partial\left(  \partial_{\nu}%
Q^{A}\right)  }T_{(c),\text{ }B}^{\text{ \ \ }A}Q^{B}-\frac{\partial L_{0}%
}{\partial\left(  \partial_{\nu}\partial_{\rho}A_{~\mu}^{a}\right)
}f_{c\text{ }b}^{\text{ }a}\partial_{\rho}A_{\text{ }\mu}^{b}+\\
&  +\left[  \partial_{\rho}\frac{\partial L_{0}}{\partial\left(
\partial_{\rho}\partial_{\nu}A_{~\mu}^{a}\right)  }-\frac{\partial L_{0}%
}{\partial\left(  \partial_{\nu}A_{~\mu}^{a}\right)  }\right]  f_{c\text{ }%
b}^{\text{ }a}A_{\text{ }\mu}^{b}+\partial_{\mu}\frac{\partial L_{0}}%
{\partial\left(  \partial_{(\mu}A_{~\nu)}^{c}\right)  }%
\end{align*}
and%
\begin{align*}
\bar{J}_{c}^{~\nu} &  =-\frac{\partial L_{T}}{\partial\left(
\partial_{\nu
}Q^{A}\right)  }T_{(c),\text{ }B}^{\text{ \ \ }A}Q^{B}-\frac{\partial L_{0}%
}{\partial\left(  \partial_{\nu}\partial_{\rho}A_{~\mu}^{a}\right)
}f_{c\text{ }b}^{\text{ }a}\partial_{\rho}A_{\text{ }\mu}^{b}+\\
&  +\left[  \partial_{\rho}\frac{\partial L_{0}}{\partial\left(
\partial_{\rho}\partial_{\nu}A_{~\mu}^{a}\right)  }-\frac{\partial L_{0}%
}{\partial\left(  \partial_{\nu}A_{~\mu}^{a}\right)  }\right]  f_{c\text{ }%
b}^{\text{ }a}A_{\text{ }\mu}^{b}-\partial_{\mu}\frac{\partial L_{0}}%
{\partial\left(  \partial_{\lbrack\mu}A_{~\nu]}^{c}\right)  },
\end{align*}
which are directly comparable with the Noether's second order current,%
\begin{align*}
\left(  J_{N}\right)  _{c}^{~\nu} &  =-\frac{\partial
L_{T}}{\partial\left(
\partial_{\nu}Q^{A}\right)  }T_{(c),\text{ }B}^{\text{ \ \ }A}Q^{B}%
-\frac{\partial L_{0}}{\partial\left(
\partial_{\nu}\partial_{\rho}A_{~\mu
}^{a}\right)  }f_{c~b}^{~a}\partial_{\rho}A_{~\mu}^{b}+\\
&  +\left[  \partial_{\rho}\frac{\partial L_{0}}{\partial\left(
\partial_{\rho}\partial_{\nu}A_{~\mu}^{a}\right)  }-\frac{\partial L_{0}%
}{\partial\left(  \partial_{\nu}A_{~\mu}^{a}\right)  }\right]  f_{c~b}%
^{~a}A_{~\mu}^{b}.
\end{align*}
Then, one easily sees that the expressions of $\bar{J}$\ and
$J_{N}$\ differ between them only by the presence of topological
terms, while in $J$ there is a quasiconservation term,
$\partial_{\mu}\frac{\partial L_{0}}{\partial\left(
\partial_{(\mu}A_{~\nu)}^{c}\right)  }$, which does not contribute in the
first order case. Therefore, in the first order approach, Utiyama
and Noether's currents coincide, but in the second order case it is
necessary to perform a generalization in order to accomplish the
conservation, giving rise to the presence of topological currents.
These topological factors are naturally introduced by the presence
of the additional term $\partial_{\mu }\frac{\partial
L_{T}}{\partial\left(  \partial_{\mu}A_{~\nu}^{c}\right)  }$ in
$\bar{J}_{c}^{~\nu}$, which can be interpreted as a flux of gauge
potential, extending the usual definition of current as a variation
of \textquotedblleft energy\textquotedblright\ $L_{T}$\ with respect
to the external field $A_{~\nu}^{c}$, $\frac{\partial
L_{T}}{\partial A_{~\nu}^{c}}$.

Another way to see how to implement this generalization is to
remember that the conservation law of the current is intimately
related to the last of the
hierarchical equations. For instance, in the first order approach,%
\[
J_{c}^{~\nu}=\frac{\partial L_{T}}{\partial A_{~\nu}^{c}}=\partial_{\mu}%
\frac{\partial L_{0}}{\partial\left(
\partial_{\mu}A_{~\nu}^{c}\right)  },
\]
where we have used the equations of motion, and from where follows
the
conservation as a direct consequence of the equation%
\[
\frac{\partial L_{0}}{\partial\left(
\partial_{\mu}A_{~\nu}^{c}\right) }+\frac{\partial
L_{0}}{\partial\left(  \partial_{\nu}A_{~\mu}^{c}\right) }\equiv0.
\]
Analogously, in the second order case we extract from the equation of motion,%
\[
\bar{J}_{c}^{~\nu}=\frac{\partial L_{T}}{\partial
A_{~\nu}^{c}}-\partial_{\mu }\frac{\partial L_{0}}{\partial\left(
\partial_{\mu}A_{~\nu}^{c}\right)
}=-\partial_{\rho}\partial_{\mu}\frac{\partial L_{0}}{\partial\left(
\partial_{\rho}\partial_{\mu}A_{~\nu}^{c}\right)  },
\]
which again is conserved by virtue of the equation (\ref{1.17
linha}).

On the other hand, Noether's procedure for the calculation of the
current imply exactly the first of our hierarchical equations,
(\ref{condicao1}), giving the Noether's current in terms of $F$ and
$G$\ as just the $N_{c}^{\,\nu}-\frac{\delta L_{T}}{\delta
A_{~\nu}^{c}}$\ quantity. We observe again that this, up to
topological terms and a global signal, is the conserved current
$\bar{J}_{c}^{\,\nu}$. Therefore, we see that the Utiyama's
systematic method leads to the expected result and gives the bonus
of finding topological currents. We conclude that far from being
arbitrary, this topological current is \emph{induced} by the
structure of the hierarchical equations, arising from the local
gauge invariance principle and has direct \emph{observable}
consequences for the charges. This will be illustrated in the next
section.


\section{Applications}

\subsection{$U\left(  1\right)  $ Group: Podolsky's Generalized
Electrodynamics \label{secao-U(1)}}

Usually in the literature \cite{Frenkel} Podolsky's electrodynamics
is supposed to be the simplest generalization of Maxwell theory
whose Lagrangian, containing second-order derivatives of the
electromagnetic potentials, is gauge and Lorentz invariant and still
leads to linear local field equations. In fact, here we will show
that the second order gauge theory is able to prove that Podolsky's
electrodynamics is the unique theory which has such properties.

Podolsky's second order theory for the electromagnetism consider the
Lagrangian density \cite{Podolsky}%
\begin{equation}
L_{0}=-\frac{1}{4}F_{\mu\nu}F^{\mu\nu}+\frac{a^{2}}{2}\partial^{\nu}F_{\mu\nu
}\partial_{\beta}F^{\mu\beta}. \label{LPodolsky}%
\end{equation}

We will check that the Podolsky's theory fulfils the condition for a
second order gauge theory. The equations to be satisfied are
(\ref{1.15}) to (\ref{1.17 linha}). The first equation is
automatically verified for the $U\left(  1\right)  $ group by virtue
of the nullification of the structure constants. Therefrom, for this
group any scalar constructed with the tensors $F$ and $G$ is a
possible Lagrangian, in principle.

The next two equations are identical to the first order ones%
\begin{align*}
\frac{\partial L_{0}}{\partial A_{\nu}}  &  \equiv0,\\
\frac{\partial L_{0}}{\partial\left(  \partial_{\nu}A_{\alpha}\right)  }%
+\frac{\partial L_{0}}{\partial\left(
\partial_{\alpha}A_{\nu}\right)  }  & \equiv0,
\end{align*}
and consenquently hold in the case of Podolsky Lagrangian
(\ref{LPodolsky}).

It remains to be verified the last of the hierarchical equations.
From
(\ref{LPodolsky}),%
\[
\frac{\partial L_{0}}{\partial\left(
\partial_{\rho}\partial_{\sigma }A_{\text{ }\tau}\right)
}=a^{2}\eta^{\tau\rho}\left(  \partial^{\sigma
}\partial^{\varepsilon}A_{\text{ }\varepsilon}-\partial^{\varepsilon}%
\partial_{\varepsilon}A^{\sigma}\right)  -\tau\leftrightarrow\sigma,
\]
and by cyclic permutation,%
\[
\frac{\partial L_{0}}{\partial\left(
\partial_{\rho}\partial_{\sigma}A_{\tau
}\right)  }+\frac{\partial L_{0}}{\partial\left(  \partial_{\sigma}%
\partial_{\tau}A_{\rho}\right)  }+\frac{\partial L_{0}}{\partial\left(
\partial_{\tau}\partial_{\rho}A_{\sigma}\right)  }=0
\]
and we see that the Podolsky's theory is in fact a second order
gauge theory \emph{a la }Utiyama.

Alternatively, one can start straight ahead from the second order
gauge theory
for the $U\left(  1\right)  $ group and write the $G$ tensor (\ref{Gnovo}),%
\[
G_{\beta\rho\sigma}=\partial_{\beta}F_{\rho\sigma},
\]
from which we can construct a vector, e.g.,%
\[
G_{\,\rho\beta}^{\beta}=\partial^{\beta}F_{\rho\beta}=\mathcal{G}_{\rho}%
\]
and obtain a second order Lagrangian equivalent to the additional
term
proposed by Podolsky:%
\[
L_{0}^{P}=\frac{a^{2}}{2}\mathcal{G}^{\rho}\mathcal{G}_{\rho}=\frac{a^{2}}%
{2}\partial^{\beta}F_{\rho\beta}\partial_{\lambda}F^{\rho\lambda}.
\]

Notwithstanding, there are other possible Lagrangians densities. For
instance, consider\footnote{Of course the \emph{complete} Lagrangian
for the gauge field
is the Maxwell one \emph{plus} the term under discussion.}%
\[
L_{G}=b^{2}G^{\beta\rho\sigma}G_{\beta\rho\sigma}=b^{2}\partial^{\beta}%
F^{\rho\sigma}\partial_{\beta}F_{\rho\sigma},
\]
which satisfies all the hierarchical equations by the same arguments
given before.

Using the Bianchi identity (\ref{Bianchi}) one concludes%
\[
G^{\beta\rho\sigma}G_{\beta\rho\sigma}=2\partial^{\beta}F^{\rho\sigma}%
\partial_{\sigma}F_{\rho\beta}.
\]
Perceive that this is the Podolsky Lagrangian apart from a surface term:%
\begin{align*}
\partial^{\beta}F^{\rho\sigma}\partial_{\sigma}F_{\rho\beta}  &
=\partial_{\sigma}\left(
\partial^{\beta}F^{\rho\sigma}F_{\rho\beta}\right)
-\partial^{\beta}\left(
\partial_{\sigma}F^{\rho\sigma}F_{\rho\beta}\right)
+\partial_{\sigma}F^{\rho\sigma}\partial^{\beta}F_{\rho\beta}.
\end{align*}
Then $L_{G}$ is equivalent to Podolsky's taking
$2b^{2}=\frac{a^{2}}{2}$.

We can explore another order of contraction to the indices of $G$:%
\[
\tilde{L}_{G}=c^{2}G^{\beta\rho\sigma}G_{\rho\beta\sigma}=c^{2}\partial
^{\beta}F^{\sigma\rho}\partial_{\rho}F_{\sigma\beta}.
\]

By the cyclicity symmetry (\ref{Bianchi})%
\[
\tilde{L}_{G}=c^{2}\partial^{\beta}F^{\sigma\rho}\partial_{\rho}F_{\sigma
\beta}.
\]
Again, this is Podolsky Lagrangian, except by a surface term, with
$c^{2}=\frac{a^{2}}{2}$.

Therefore we can infer that \emph{all} the quadratic Lagrangian in
$G$ are reducible to the Podolsky's form since there are only two
possible cases for indices contractions with respect to the
derivative of $F$: or the indices in the derivatives are contracted
between them or they are contracted with one of the $FF$. In the
first case one uses the Bianchi identity, which reduce all
contractions to the second case. But the second case is just
Podolsky's, or can be put in this form by an adequate surface term.

For these reasons, we see that the most general Lagrangian for the
$U\left(
1\right)  $ group in a theory \emph{a la} Utiyama is%
\[
L_{0}=-\frac{1}{4}F^{\mu\nu}F_{\mu\nu}+\frac{a^{2}}{2}\partial^{\beta}%
F_{\rho\beta}\partial_{\lambda}F^{\rho\lambda},
\]
which is the Podolsky Electrodynamics Lagrangian.

This proves that Podolsky Lagrangian is the unique linear second
order generalization from Maxwell theory compatible with the gauge
principle.

\subsection{$U\left(  1\right)  $ Currents and the Mass of the
Photon\label{MassPhoton}}

With (\ref{corrente}) for the case $U\left(  1\right)  $,%
\[
J^{\nu}=j^{\nu}=-\frac{\partial L_{T}}{\partial\left(  \nabla_{\nu}%
Q^{A}\right)  }T_{\,B}^{A}Q^{B},
\]
and this means that the second order formalism has no effect on the
current if the symmetry is $U\left(  1\right)  $.

Now if we employ (\ref{corrente2}),%
\[
\bar{J}^{\nu}=j^{\nu}-\partial_{\mu}\frac{\partial L_{0}}{\partial
F_{\left[ \mu\nu\right]  }}=j^{\nu}-\partial_{\mu}F^{\nu\mu},
\]
which differs from $J^{\nu}$\ by a topological term.

The equation of motion, in vacuum, for Podolsky's electrodynamics is%
\[
\left(  1+a^{2}\square\right)
\partial_{\lambda}F^{\mu\lambda}\left( x\right)  =0,
\]
which, with the Lorentz gauge condition, $\partial_{\mu}A^{\mu}=0$, reduces to%
\begin{equation}
\left(  1+a^{2}\square\right)  \square A^{\mu}\left(  x\right)  =0.
\label{eq motion}%
\end{equation}
Extracting the Fourier transform, we find two possible dispersion relations:%
\begin{align*}
p^{2}  &  =0,\\
p^{2}-\frac{1}{a^{2}}  &  =0.
\end{align*}
The first one corresponds to a massless mode,
$E^{2}-\mathbf{p}^{2}=0$, while
the second is a massive solution:%
\begin{equation}
E^{2}=\mathbf{p}^{2}+\frac{1}{a^{2}}~,\;m=\frac{1}{a}. \label{photon mass}%
\end{equation}

In the case of a static isolated point charged particle, equation
(\ref{eq motion}) gives:%
\begin{gather*}
-\left(  1-a^{2}\nabla^{2}\right)  \nabla^{2}\varphi\left(
\mathbf{r}\right)
=4\pi\rho\left(  \mathbf{r}\right)  ,\\
\rho\left(  \mathbf{r}\right)  =-e\delta\left(  \mathbf{r}\right)  .
\end{gather*}
We have selected a Lorentz gauge such that $A^{\mu}=\left(  \varphi
,\mathbf{0}\right)  $, and used the signature $\eta=diag\left(
1,-1,-1,-1\right)  $. Via Fourier transform, we find%
\[
\varphi\left(  \mathbf{r}\right)  =-\frac{e}{r}\left(  1-e^{-\frac{r}{a}%
}\right)  .
\]
The finite reach of the massive mode suggests the presence of a
shielded region around the point particle.

The conserved current is given by%
\[
\bar{J}^{0}\left(  \mathbf{r}\right)  =-e\left(  \delta\left(  \mathbf{r}%
\right)  +\frac{1}{a^{2}r}e^{-\frac{r}{a}}\right)
\;,\;\bar{J}^{k}\equiv0.
\]
Integrating this density over the whole space, we obtain the conserved charge,%
\[
q=-e\left(  1-4\pi\right)
\]
which is just a simple renormalization of the naked charge. However,
following Moniz and Sharp \cite{MonizSharp}, the quantum theory of
the nonrelativistic electron provides a natural cutoff at short
distances whose magnitude is of the order of the Compton length
$\varepsilon$. Employing this idea of a
shielded region, the effective conserved charge obtained is%
\begin{equation}
q_{eff}=-e\left(  1-4\pi\left[  \frac{a+\varepsilon}{a}\exp\left(
-\frac{\varepsilon}{a}\right)  \right]  \right)  . \label{Qeff}%
\end{equation}

This effective charge can be applied to estimate the mass of the
photon,
according to (\ref{photon mass}). Equation (\ref{Qeff}) can be rewritten as%
\[
\sigma_{e}^{rel}\equiv\frac{q_{eff}-e}{e}=4\pi\frac{a+\varepsilon}{a}%
\exp\left(  -\frac{\varepsilon}{a}\right)  .
\]
This quantity can be interpreted as the relative uncertainty of the
electron charge. Solving this transcendental equation for Podolsky's
parameter $a$, we
find a measure of mass for the massive photon,%
\[
m=\frac{\hbar}{ac}~.
\]
With experimental data from \cite{PDG} and a simple propagation of
errors
neglecting the uncertainty in $\sigma_{e}^{rel}$, we obtain%
\begin{align*}
a  &  =1.105\,868\,617\left(  14\right)  \times10^{-11}\,cm~,\\
m  &  =1.985\,370\,21\left(  17\right)  \times10^{-13}\,eV~.
\end{align*}
This estimative cannot be ruled out in accordance with experimental
data reported in \cite{PDG}\ and \cite{Ryan} considering null tests
of Coulomb law. These references give $m<\left(  9.0\pm8.1\right)
\times10^{-10}~eV$. In the context of an effective classical theory,
this mass must be interpreted as an energy scale\ characterizing the
regime where the Podolsky's electrodynamics becomes relevant, in a
similar way as the phonon mass characterizes the effects of particle
excitations in a cristal \cite{Phonon}. It is meaningless to think
of a rest mass for a phonon, since it is an excitation of the
vibrations in a cristal. The same interpretation can be applied for
the photon in Podolsky's theory.

We emphasize that these results were derived using a completely
classical approach and not following a construction of effective
classical theories from their quantum versions. This is the reason
why the renormalized charge is found and no estimative for the
photon mass can be made in the absence of a shielded region. Even
taking a shielded region, if Noether's current was considered, the
effective charge would be just $e$ and, again, no mass for the
photon could be found, as well no renormalization for the charge
would be obtained.

\subsection{$SU\left(  N\right)  $ Group: The Lagrangian
AAB\label{AlekseevArbuzov}}

As a non-abelian example we apply the second order gauge theory to
the description of Alekseev-Arbuzov-Baikov's effective Lagrangian
\cite{Baikov} proposed to eliminate infrared divergences in $SU(N)$
theories. In our
notation, it is%
\[
L_{eff}=\frac{1}{4M^{2}}G_{\lambda\mu\nu}^{a}G_{a}^{~\lambda\mu\nu}+\frac
{1}{6M^{2}}f_{b~c}^{~a}F^{b\lambda\mu}F_{a\mu}^{\;\;\tau}F_{~\tau\lambda}^{c}%
\]
where $M$ is the mass scale of the infrared gluon. This scalar
density satisfies the condition (\ref{1.15Linha}), a fact verified
after using the total antisymmetry of the structure constants.

The conserved current for this theory can be obtained from (\ref{corrente2}):%
\begin{align*}
\left(  \bar{J}_{eff}\right)  _{c}^{\,\nu} &  =\frac{1}{M^{2}}f_{c\,b}%
^{\,a}\left(  A_{\,\mu}^{b}\left[  D_{a~\rho}^{~d}G_{d}^{~\rho\nu\mu}%
+f_{a\;e}^{~d}F^{e\mu\tau}F_{d\tau}^{~\;\;\nu}\right]  -\frac{1}{2}%
F_{\,\rho\mu}^{b}G_{a}^{~\nu\rho\mu}\right)  +\\
&  +\frac{1}{M^{2}}f_{c\text{ }b}^{\text{ }d}\partial_{\mu}\left(
A_{\text{
}\rho}^{b}G_{d}^{~\rho\nu\mu}+F^{b\mu\tau}F_{d\tau}^{~\;\;\nu}\right)
\end{align*}
which differs from the current in \cite{Confinamento}\ only by
topological terms.

The equivalence of 1-loop infrared regime and the second order gauge
theory shows that an analysis of high order gauge theories could
give more information about higher loop expansions.


\section{General Remarks}

In this work we saw how Utiyama's approach to gauge theory can be
extended to include second order derivatives of the gauge potential.
The development showed the need for the introduction of a new object
$G_{\text{ }\beta \rho\sigma}^{c}$, as well as the already familiar
field strength $F_{\,\beta\sigma}^{e}$. Both are covariant and have
a Bianchi identity associated with it, which was derived from
hierarchical equations for the auxiliary field Lagrangian.

The nullification of the variation of the total Lagrangian conducted
us to conserved currents that have two possible definitions. The
first one was defined as in the first order treatment, but it is
only \emph{quasi}conserved. Conversely, we could define a new
current, which differs structurally from the one proposed by
Utiyama, but is characterized by its conservation. Another advantage
of the general Utiyama's procedure is the obtainment of
\emph{topological} currents from a \emph{local} invariance. It is a
new tool to study topological aspects of gauge theories from their
local symmetry, and must be better investigated in the future. These
topological currents are not arbitrary, but rather \emph{implied} by
the very structure of the field and hierarchical equations, as will
be better discussed in the section \ref{HighOrderGen}.

We also saw how the second order development can be implemented in
the case of the $U\left(  1\right)  $\ group leading to the
well-known Podolsky Generalized Electrodynamics, whose Lagrangian
was demonstrated to be the unique possible extension up to surface
terms. As an example for the more complicated case of a non-abelian
theory, we applied the second order gauge theory to the
Alekssev-Arbuzov-Baikov effective Lagrangian valid in the infrared
regime, demonstrating that the conserved current (\ref{corrente2})
and the one obtained in \cite{Confinamento} differ by surface terms.
In this way, we have shown that 1-loop effective terms in the action
can be found from first-principle calculations imposing the gauge
symmetry to second order Lagrangians. In addition, an extension of
Utiyama's method to higher orders derivatives could give important
information about higher loop expansions.

The conserved current laws were used to estimate the energy scale of
the photon massive mode, in the electrostatic regime, and the result
is in accordance with the present observed limits. This massive mode
was engendered by the topological terms in the conserved current
(\ref{corrente2}) in accordance with the known fact that topological
terms are responsible for dynamic mass generation mechanisms
\cite{Jackiw}. Besides, the conserved current $\bar{J}$ gives the
\emph{renormalized charge}, which is quite natural in view of the
equivalence among second order gauge theory and 1-loop expansions.
On the other hand, the Noether's current only gives bare charges.
Notice that in the $U\left(  1\right)  $ case both $J$ and $\bar{J}$
are conserved.

Among the perspectives opened by this work we can cite another
natural extension to Utiyama's work allowing the Lagrangian to
depend also on the second derivative of the matter field
$\partial^{2}Q$, which is presently under study. Other possibilities
are commented in the next sections.

\subsection{Generalization for Higher Order Theories\label{HighOrderGen}}

In order to get a generalization to higher orders, we note that some
structures apparently repeat themselves in the first and second
order formalisms. For example, the introduction of the gauge
potential as a
compensator to achieve the local invariance and the hypothesis $L_{0}%
=L_{0}\left(  A,\partial A\right)  $ leads to\ conclude that the
derivative of
$A$ must respect the symmetry%
\begin{equation}
\frac{\partial L_{0}}{\partial\left(
\partial_{\mu}A_{\,\nu}^{a}\right) }+\frac{\partial
L_{0}}{\partial\left(  \partial_{\nu}A_{\,\mu}^{a}\right)
}\equiv0. \label{ultima 1 ord}%
\end{equation}
In the second order extension one finds that the second derivative
accomplishes%
\begin{equation}
\frac{\partial L_{0}}{\partial\left(
\partial_{\alpha}\partial_{\nu}A_{\text{ }\mu}^{a}\right)
}+\frac{\partial L_{0}}{\partial\left(  \partial_{\nu
}\partial_{\mu}A_{\text{ }\alpha}^{a}\right)  }+\frac{\partial L_{0}}%
{\partial\left(  \partial_{\mu}\partial_{\alpha}A_{\text{
}\nu}^{a}\right)
}\equiv0. \label{ultima 2 ord}%
\end{equation}

Therefore, it seems natural to suppose for a $n$th-order theory the
higher
order derivative appearing only in an object%
\[
R_{\,\alpha_{1}...\alpha_{n}}^{a}\equiv\sum_{P\left\{  \alpha_{1}%
,...,\alpha_{n}\right\}  }C^{\left\{
\alpha_{1},...,\alpha_{n}\right\}
}\partial_{\alpha_{1}}...\partial_{\alpha_{n-1}}A_{\,\alpha_{n}}^{a}%
\]
carrying a cyclic permutation symmetry,%
\[
\sum_{P_{cycl}\left\{  \alpha_{1},...,\alpha_{n}\right\}
}R_{\,\alpha _{1}...\alpha_{n}}^{a}\equiv0,
\]
where $P\left\{  \alpha_{1},...,\alpha_{n}\right\}  $\ is to denote
the permutation of indices and $P_{cycl}\left\{
\alpha_{1},...,\alpha _{n}\right\}  $\ the cyclic one. This identity
might be understood as a generalization of the Bianchi identity.

Other recurrent structures are as follows. Introduced the auxiliary
field $A$,
the tensor $F$ can be written as (\ref{Fnovo})%
\[
F_{\,\beta\sigma}^{e}=D_{\,f[\beta}^{e}A_{\,\sigma]}^{f}%
\]
In the second order case a new object comes into place (\ref{Gnovissimo})%
\[
G_{\text{ }\beta\rho\sigma}^{a}=D_{\,c[\beta}^{a}F_{\,\rho]\sigma}^{c}%
\]
In time, we may conjecture for a third order theory the emerging of
a new
object whose functional form is%
\[
H_{\,\alpha\beta\rho\sigma}^{a}=D_{\,c[\mu}^{a}G_{\text{
}\beta]\rho\sigma }^{c},
\]
respecting, as said before, a Bianchi identity%
\[
H_{\,\mu\beta\rho\sigma}^{a}+H_{\,\beta\rho\sigma\mu}^{a}+H_{\,\rho\sigma
\mu\beta}^{\alpha}+H_{\,\sigma\mu\beta\rho}^{a}=0,
\]
Remembering that for a general Lie group vector,
\[
D_{\,c[\mu}^{a}G_{\text{ }\beta]\rho\sigma}^{c}=\left[
D_{\mu},D_{\beta
}\right]  _{\,\ d}^{a}F_{\,\rho\sigma}^{d}=f_{d\text{ }e}^{\text{ }a}%
F_{\,\mu\beta}^{e}F_{\,\rho\sigma}^{d},
\]
one finds%
\begin{align*}
&
H_{\,\mu\beta\rho\sigma}^{a}+H_{\,\beta\rho\sigma\mu}^{a}+H_{\,\rho\sigma
\mu\beta}^{\alpha}+H_{\,\sigma\mu\beta\rho}^{a}=\\
& =D_{\,c[\mu}^{a}G_{\text{
}\beta]\rho\sigma}^{c}+D_{\,c[\beta}^{a}G_{\text{
}\rho]\sigma\mu}^{c}+D_{\,c[\rho}^{a}G_{\text{ }\sigma]\mu\beta}%
^{c}+D_{\,c[\sigma}^{a}G_{\text{ }\mu]\beta\rho}^{c}=\\
& =f_{d\text{ }e}^{\text{
}a}F_{\,\mu\beta}^{e}F_{\,\rho\sigma}^{d}+f_{d\text{ }e}^{\text{
}a}F_{\,\beta\rho}^{e}F_{\,\sigma\mu}^{d}-f_{d\text{ }e}^{\text{
}a}F_{\,\rho\sigma}^{d}F_{\,\mu\beta}^{e}-f_{d\text{ }e}^{\text{
}a}F_{\,\sigma\mu}^{d}F_{\,\beta\rho}^{e}=0.
\end{align*}
and the structure is in fact consistent. It is important to note
that this is not a rigorous proof of the third order structure, but
a significant indication of its form.

We have seen that the local gauge invariance condition on the first
order
Lagrangian is \cite{Utiyama}%
\[
\frac{\partial L_{0}}{\partial F_{\,\rho\sigma}^{d}}f_{c\,h}^{\,d}%
F_{\rho\sigma}^{h}\equiv0
\]
and we have proved for the second order theory the generalization%
\[
\frac{\partial L_{0}}{\partial F_{\,\rho\sigma}^{d}}f_{c\,h}^{\,d}%
F_{\rho\sigma}^{h}+\frac{\partial L_{0}}{\partial G_{\text{
}\beta\rho\sigma }^{d}}f_{c\text{ }h}^{\text{ }d}G_{\text{
}\beta\rho\sigma}^{h}\equiv0
\]
Therefore, it appears natural to expect the restriction
\[
\frac{\partial L_{0}}{\partial F_{\,\rho\sigma}^{d}}f_{c\,h}^{\,d}%
F_{\rho\sigma}^{h}+\frac{\partial L_{0}}{\partial G_{\text{
}\beta\rho\sigma }^{d}}f_{c\text{ }h}^{\text{ }d}G_{\text{
}\beta\rho\sigma}^{h}+\frac{\partial L_{0}}{\partial H_{\text{
}\alpha\beta\rho\sigma}^{d}}f_{c\text{ }h}^{\text{ }d}H_{\text{
}\alpha\beta\rho\sigma}^{h}\equiv0
\]
on the third order Lagrangian of the auxiliary field.

The structures identified until now indicate that for still higher
order theories one can hope for objects given in terms of
Fock-Ivanenko derivatives of the quantities of the former order.
Another feature is that the higher order objects also are supposed
to satisfy a Bianchi-like identity.

Regarding conservation laws, Utiyama defined the current%
\[
J_{a}^{\,\mu}\equiv\frac{\partial L_{T}}{\partial A_{\,\mu}^{a}}%
\]
which is a conserved quantity for the first order theory,
$L_{0}=L_{0}\left(
A,\partial A\right)  $, under the equation of motion%
\[
\frac{\delta L_{0}}{\delta A_{\,\mu}^{a}}=\frac{\partial
L_{T}}{\partial A_{\,\mu}^{a}}-\partial_{\nu}\frac{\partial
L_{T}}{\partial\left(
\partial_{\nu}A_{\,\mu}^{a}\right)  }=0.
\]
We have seen that this definition of current is not appropriate for
the second order approach, $L_{0}=L_{0}\left(  A,\partial
A,\partial^{2}A\right)  $, since it is not a conserved object. To
solve this problem we defined
\[
\bar{J}_{a}^{\,\mu}\equiv\frac{\partial L_{T}}{\partial A_{\,\mu}^{a}%
}-\partial_{\nu}\frac{\partial L_{T}}{\partial\left(
\partial_{\nu}A_{\,\mu
}^{a}\right)  }%
\]
as the expression of the conserved current valid on mass shell,
which in this
case reads%
\[
\frac{\delta L_{0}}{\delta A_{\,\mu}^{a}}=\frac{\partial
L_{T}}{\partial A_{\,\mu}^{a}}-\partial_{\nu}\frac{\partial
L_{T}}{\partial\left(
\partial_{\nu}A_{\,\mu}^{a}\right)  }+\partial_{\rho}\partial_{\nu}%
\frac{\partial L_{T}}{\partial\left(
\partial_{\rho}\partial_{\nu}A_{\,\mu }^{a}\right)  }=0.
\]
Notice that in each case we have just isolated the last term in the
Euler-Lagrange equation and defined the remaining terms as the
conserved current. The conservation is, in fact, assured by the
symmetry imposed by the last of the hierarchical equations in each
case, namely (\ref{ultima 1 ord}) and (\ref{ultima 2 ord})
respectively.

This reasoning may be applied to extended theories: in a $n$-order
theory $L_{0}=L_{0}\left(  A,\partial A,...,\partial^{n}A\right)  $
the current may be defined as the first $n$ terms of the
Euler-Lagrange equation. The currents constructed following this
recipe would differ from the Noether's one by topological terms, a
fact that reflects upon the values of the charges. Which values is
the correct one stands to be selected by experience.

\subsection{Geometrical Aspects}

Geometrical aspects are not evident in the Utiyama's approach since
it is essentially an algebraic implementation of the local symmetry
principle. A geometrical interpretation of $G$ could help one to
understand the recurrent emergence of the Fock-Ivanenko derivative
and the general structure of higher orders Lagrangians outlined
above.

Moreover, the comprehension of the geometry seems to be fundamental
if one wants to describe \emph{higher order gravitation}\ as a
direct application of Utiyama's second order procedure, a work now
under construction by the authors. This could also\ illuminate some
features of the nonrenormalizability problem in gravitation.

\subsection{Application to QCD}

Besides the analysis of the AAB Lagrangian, it would be interesting
to perform an 1-loop calculation and check if the cancelling present
in Podolsky Electrodynamics \cite{Podolsky}\ is also manifested in
QCD.

However, one must consider which quadratic form in $G$ will be
explored. Due to the nonlinearity of the theory, it is possible that
\emph{all} combinations be necessary in order to obtain a
renormalizable theory. Another probable feature is the impossibility
of closing the number of derivatives (counterterms), as in the case
of gravitational field. Some tentative works in this direction were
undertaken in \cite{Confinamento}.

An additional point is to investigate if higher order terms gives
more information about the confinement phenomenon using, for
example, Wilson criterion, as noted in \cite{Confinamento}. Once
more, to recognize the geometrical connection which defines the
holonomies would be a good guide.

\subsection{Constraint Analysis}

The local gauge symmetry imply the existence of constraints in any
order of derivatives which require a special attention in order to
construct the Hamiltonian description.

The analysis of the constraints for a second order Lagrangian can be
implemented using several approaches, such as\ that in
\cite{PimentaGalvao} for Podolsky Electrodynamics. Nevertheless, the
study of an specific Lagrangian for non-abelian groups is
interesting in view of the applications for QCD and gravitation.
This study is presently under implementation by means of
Hamilton-Jacobi technique \cite{HamiltonJacobi}.


\begin{center}
ACKNOWLEDGEMENTS
\end{center}

R. R. C. and C. A. M. M. are grateful to Funda\c{c}\~{a}o de Amparo
\`{a} Pesquisa do Estado de S. Paulo (FAPESP) for support (grants
02/05763-8 and 01/12584-0 respectively) and P. J. P. thanks to CTA
staff for the incentive. The authors acknowledge Prof. B. M.
Pimentel for introducing them to the original Utiyama's work, Prof.
R. Aldrovandi for the careful reading of the original manuscript and
useful suggestions and Prof. J. A. Accioly for the enthusiastic
incentive to this project.



\begin{thebibliography}{99}                                                                                               %


\bibitem {Utiyama}R. Utiyama, Phys. Rev.\textit{\ }\textbf{101}, 1597 (1956).

\bibitem {O-P}V. I. Ogievetski and I. V. Polubarinov, Nuovo Cim.\textit{\ }%
\textbf{23}, 173 (1962).

\bibitem {NakaLivro}N. Nakanishi and I. Ojima, \textit{Covariant Operator
Formalism of Gauge Theories and Quantum Gravity}, World Scientific,
Lecture Notes in Physics, Vol.27 (1991).

\bibitem {Bfield}C. A. M. de Melo, B. M. Pimentel \& P. J. Pompeia -
\textit{Schwinger's Principle and Gauge Fixing in the Free
Electromagnetic Field}, [hep-th/0502229]; To appear in \textit{Il
Nuovo Cimento }\textbf{B}.

\bibitem {Poincare}H. Poincar\'{e}, \textit{Science and Hypothesis}, Dover, (1952).

\bibitem {Ostrogradsky}M. Ostrogradsky, Mem. Acad. St. Petersburg\textit{\ }%
\textbf{6} (4), 385 (1850); E. T. Whitaker, \textit{Treatise on the
Analytical Dynamics of Particles and Rigid Bodies}, Cambridge
University Press, 4th ed. (1959).

\bibitem {Bopp}F. Bopp, Ann. Physik\textit{\ }\textbf{38}, 345 (1940).

\bibitem {Podolsky}B. Podolsky, Phys. Rev.\textit{\ }\textbf{62}, 68 (1942);
B. Podolsky and C. Kikuchi, \textit{ibid }\textbf{65}, 228 (1944);
B. Podolsky and P. Schwed, Rev. Mod. Phys.\textit{\ }\textbf{20}, 40
(1948).

\bibitem {Frenkel}J. Frenkel, Phys. Rev. E\textit{\ }\textbf{54}, 5859 (1996);
J. Frenkel and R. B. Santos, Int. J. Mod. Phys. B\textit{\
}\textbf{13}, 315 (1999).

\bibitem {Green}A. Green, Phys. Rev.\textit{\ }\textbf{73}, 26 (1948);
\textbf{75}, 1926 (1949).

\bibitem {Confinamento}A. I. Alekseev and B. A. Arbuzov, Theor. Math. Phys.
\textbf{59}, 372 (1984); M. Baker, J. S. Ball and F. Zachariasen,
Nucl. Phys. B\textit{\ }\textbf{229}, 445 (1983).

\bibitem {Fock}V. A. Fock, Z. Phys. \textbf{57}, 261 (1929).

\bibitem {Felsager}B. Felsager - \textit{Geometry, Particles, and Fields},
Springer-Verlag, New York, (2000).

\bibitem {MonizSharp}E. J. Moniz and D. H. Sharp, Phys. Rev. D \textbf{10},
1133 (1974); \textbf{15}, 2850 (1977).

\bibitem {PDG}S. Eidelman \textit{et al.}, Phys. Lett. B \textbf{592}, 1 (2004).

\bibitem {Ryan}J. J. Ryan, F. Accetta and R. H. Austin, Phys. Rev. D
\textbf{32}, 802 (1985).

\bibitem {Phonon}R. Kubo and T. Nagamiya, \textit{Solid State Physics}
McGraw-Hill, New York, (1968).

\bibitem {Baikov}A. I. Alekseev, B. A. Arbuzov and V. A. Baikov, Theor. Math.
Phys. \textbf{52}, 739 (1982).

\bibitem {Jackiw}G. Dvali, R. Jackiw, So-Young Pi, \textit{Topological Mass
Generation in Four Dimensions}, Report-n%
${{}^o}$%
: MIT-CTP-3704 [hep-th/0511175].

\bibitem {PimentaGalvao}C. A. P. Galv\~{a}o and B. M. Pimentel, Can. J. Phys.
\textbf{66}, 460 (1988).

\bibitem {HamiltonJacobi}Y. G\"{u}ler, Il Nuovo Cimento\textit{\ }%
\textbf{B107}, 1398 (1992); B. M. Pimentel, R. G. Teixeira and J. L.
Tomazelli, Ann. Phys.\textit{\ }\textbf{267}, 75 (1998).
\end{thebibliography}
\end{document}